\documentclass[11pt,aps,pra]{revtex4}
\usepackage{verbatim}
\usepackage{amsmath,amssymb,bbm}
\usepackage{amsfonts}
\usepackage{graphicx}
\usepackage{pgf}

\newcommand{\si}{\sigma}
\newcommand{\cm}{\mathrm{cm}}
\newcommand{\rt}{\mathrm{rot}}
\newcommand\ro{\hat\rho}
\newcommand\Ho{\hat H}
\newcommand\rb{\mathbf{r}}
\newcommand\rbp{\mathbf{r^\prime}}

\newcommand\xbh{\mathbf{\hat x}}
\newcommand\Xb{\mathbf{X}}
\newcommand\Xbh{\mathbf{\hat X}}
\newcommand\Yb{\mathbf{Y}}

\newcommand\pbh{\mathbf{\hat p}}
\newcommand\Pbh{\mathbf{\hat P}}
\newcommand\nb{\mathbf{n}}

\newcommand\vro{\hat \varrho}
\newcommand\Dcal{\mathcal{D}}
\newcommand\dd{\mathrm{d}}
\newcommand\kb{\mathbf{k}}

\newcommand\erm{\mathrm{e}}

\newcommand\im{\mathrm{i}}

\begin{document}

\title{Two invariant surface-tensors determine CSL of  massive body wave function}
\author{Lajos Di\'osi}
\email{diosi.lajos@wigner.mta.hu}
\affiliation{Wigner Research Center for Physics\\ 
             H-1525 Budapest 114, P.O.Box 49, Hungary}

\date{\today}

\begin{abstract}
Decoherence of massive body wave function under Continuous Spontaneous Localization is reconsidered. 
It is shown for homogeneous probes with wave functions narrow in position and angle that 
decoherence is a surface effect. Corresponding new surface integrals 
are derived as the main result. Probe's constant density and two completely 
geometric surface-dependent invariant tensors encode full dependence 
of positional and angular decoherence of masses, irrespective of their microscopic structure.
The two surface-tensors offer a new insight into CSL  and a flexible approach to design 
laboratory test masses.
\end{abstract}

\maketitle

\section{Introduction}\label{I}
\begin{figure}
\centering
    \includegraphics[width=0.50\textwidth, angle=0]{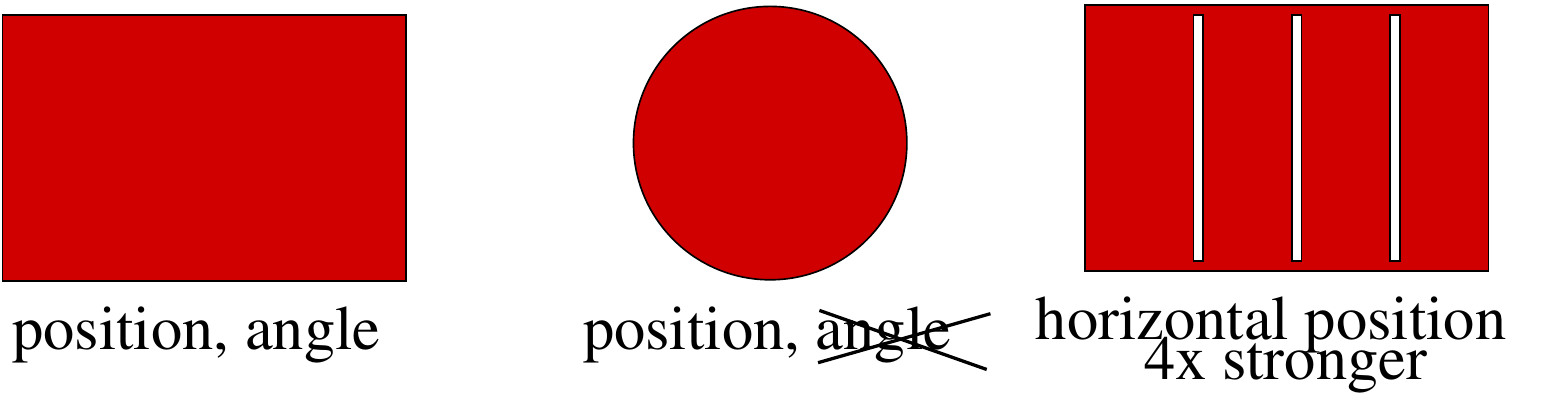}
\caption{For a generic shape, both position \& angle decohere (left). For a sphere, angle
does not decohere (middle). $N$ perpendicular gaps enhance longitudinal decoherence by
a factor about (N+1) (right).}
\label{fig-1}
\end{figure}

Spontaneous decoherence and collapse models, reviewed e.g. by  \cite{BasGhi03,Basetal13} 
share the form of modified von Neumann equation of motion for the quantum state $\ro$:  
\begin{equation}\label{ME}
\frac{d\ro}{dt}=-\frac{\im}{\hbar}[\Ho,\ro]+\Dcal\ro,
\end{equation}
where $\Ho$ is the many-body Hamiltonian of masses $m_a$ with positions $\xbh_a$ and momenta $\pbh_a$, resp., for $a=1,2,\dots$.
The term of spontaneous decoherence takes this generic form:
\begin{equation}\label{Dec0}
\Dcal\ro=-\int\int D(\rb-\rbp)[\vro(\rb),[\vro(\rbp),\ro]]\dd\rb\dd\rbp,
\end{equation}
containing the mass density operator at location $\rb$:
\begin{equation}\label{densop}
\vro(\rb)=\sum_a m_a\delta(\rb-\xbh_a).
\end{equation}
The non-negative decoherence kernel $D(\rb-\rbp)$ is model dependent.
In a conference talk \cite{Dio14}, I compared  some characteristic
features of the two leading proposals, the CSL of Ghirardi, Pearle, and  Rimini, and 
the DP-model of Penrose and myself \cite{Pen96,Dio87}. I visualized some
observations on CSL in Fig. \ref{fig-1} that have been waiting for
mathematical formulation until now. 
In recent literature, the central mathematical object is the \emph{geometric factor} of decoherence:
\begin{equation}\label{gf}
\mu_\kb=\sum_a m_a \erm^{-\im\kb\rb_a},
\end{equation}
introduced by \cite{NimHorHam14}, also
discussed by \cite{AdlBasCar19} in this volume.
This object is the Fourier-transform of the classical mass density  in the c.o.m. frame:
\begin{equation}
\mu(\rb)=\sum_a m_a \delta(\rb-\rb_a).
\end{equation}
 Usually, the contribution of the geometric factor
 is evaluated in the Fourier-representation. I am going to show that working in the physical space
 instead of Fourier's is not only possible but even desirable. 
 
 In Sec. \ref{II} we recapitulate the decoherence of c.o.m. motion in terms of the geometric factor.
For constant density probes, Sec. \ref{III} derives a new practical expression of the decoherence in
terms of a simple surface integral, the method is applied for angular (rotational) decoherence in Sec. \ref{IV}. 
Possible generalizations towards probes with unsharp edges and
for wider superpositions are outlined in Sec.  \ref{IV}, while Sec.  \ref{V} is for conclusion and outlook. 

\section{Center-of-mass decoherence}\label{II}
The CSL model introduces two universal parameters, collapse rate $\lambda=10^{-16}s^{-1}$, localization $\si=10^{-5}cm$,
and it contains the nuclear mass $m_N$.
The decoherence kernel $D(\rb-\rbp)$ is a Gaussian
whose nonlocal effect can be absorbed by a Gaussian smoothening of the mass density $\vro(\rb)$.
The key quantity is the $\si$-smoothened mass distribution operator: 
\begin{equation}\label{mdens}
\vro_\si(\rb)=\sum_a m_a G_\si(\rb-\xbh_a),
\end{equation} 
where $G_\si(\rb)$ is the central symmetric Gaussian distribution of width $\si$.
Then the decoherence term (\ref{Dec0}) becomes a single-integral:
\begin{equation}\label{Dec}
\Dcal\ro=-\frac{4\pi^{3/2}\lambda\si^3}{m_N^2}\int[\vro_\si(\rb),[\vro_\si(\rb),\ro]]\dd\rb.
\end{equation}
Inserting Eq. (\ref{mdens}), Fourier-representation yields this equivalent form:
\begin{equation}\label{DecFou}
\Dcal\ro=-\frac{\lambda\si^3}{2\pi^{2/3}m_N^2}\int\erm^{-\kb^2\si^2}\sum_{a,b}m_a m_b[\erm^{\im\kb\xbh_a},[\erm^{-\im\kb\xbh_b},\ro]]\dd\kb.
\end{equation}
We are interested in the c.o.m. dynamics of the total mass $M=\sum_a m_a$:
\begin{equation}\label{MEcom}
\frac{d\ro_\cm }{dt}=-\frac{\im}{\hbar}[\Ho_\cm ,\ro_\cm ]+\Dcal_\cm \ro_\cm ,
\end{equation}
where $\Xbh,\Pbh$ will stand for the c.o.m. coordinate and momentum.
To derive the c.o.m. decoherence term, substitute $\xbh_a=\Xbh+\rb_a$ in (\ref{DecFou}),
 where  $\rb_a$ are the equilibrium values of the constituent coordinates in the c.o.m. frame.
Then Eq. (\ref{DecFou}) reduces to the following c.o.m. decoherence term:
\begin{equation}\label{Deccom}
\Dcal_\cm \ro_\cm =-\frac{\lambda\si^3}{\pi^{3/2}m_N^2}\int\erm^{-\kb^2\si^2}\vert\mu_\kb\vert^2\left(\erm^{\im\kb\Xbh}\ro_\cm \erm^{-\im\kb\Xbh}-\ro_\cm \right)\dd\kb,
\end{equation}
where we recognize the presence of the geometric factor $\mu_\kb$.
At small quantum uncertainties, when $\vert\Delta\Xb\vert\ll\si$,  we use the momentum-diffusion equation as a good approximation:
\begin{equation}\label{Deccomxx}
\Dcal_\cm \ro_\cm =-\frac{\lambda\si^3}{2\pi^{3/2}m_N^2}\int\erm^{-\kb^2\si^2}\vert\mu_\kb\vert^2[\kb\Xbh,[\kb\Xbh,\ro_\cm ]]\dd\kb.
\end{equation}

\section{Invariant surface-tensor for c.o.m. decoherence}\label{III}
As we see, the geometric factor $\mu_\kb$ itself does not matter but its squared modulus does.
We consider the approximation (\ref{Deccomxx}) which allows for a spectacular simple geometric interpretation
of the relevant structure
\begin{equation}\label{geom1}
\int\erm^{-\kb^2\si^2}\vert\mu_\kb\vert^2(\kb\circ\kb)\dd\kb
=(2\pi)^3\int \nabla\mu_\si(\rb)\circ\nabla\mu_\si(\rb)\dd\rb.
\end{equation}
We can recognize $\mu_\si(\rb)$ as the $\sigma$-smoothened mass density in the c.o.m. frame.
This latter form becomes amazingly useful if the bulk is much larger than $\si$ and possesses constant density $\varrho$ 
when averaged over the scale of $\si$.
If, furthermore, we assume the density drops sharply from $\varrho$ to zero through the surface
then $\nabla\mu_\sigma(\rb)$ is vanishing everywhere but in about a $\sigma$-layer around the surface. Let $\nb$
stand for the normal vector of the surface at a given point $\rb$ and let $h$ be the height above the surface, then
\begin{equation}\label{profile0}
\nabla\mu_\si(\rb+h\nb)=-\varrho \nb g_\si(h),
\end{equation}
$g_\si(h)$ is the central Gaussian of width $\si$.
The volume integral can be rewritten, with good approximation, as an integral along $h$ and a subsequent surface integral: 
\begin{equation}\label{geom2}
(2\pi)^3\int \nabla\mu_\si(\rb)\circ\nabla\mu_\si(\rb)\dd\rb=(2\pi)^3\varrho^2\oint \nb\circ\nb\left(\int g_\si^2(h)\dd h\right)\dd S=\frac{(2\pi)^3\varrho^2}{2\pi^{1/2}\si}\oint(\nb\circ\nb)\dd S.
\end{equation}
Using Eqs. (\ref{geom1}) and (\ref{geom2}), the decoherence term (\ref{Deccomxx}) obtains the attractive form
\begin{equation}\label{Deccomxxgeom}
\Dcal_\cm \ro_\cm =-\frac{2\pi\lambda\si^2\varrho^2}{m_N^2}\oint[\nb\Xbh,[\nb\Xbh,\ro_\cm ]]\dd S.
\end{equation}
\emph{This is our main result.} It  shows that the c.o.m. decoherence is completly determined by the
constant density $\varrho$ and the shape of the body, through the \emph{surface-tensor}
\begin{equation}\label{shape1}
\mathrm{S}=:\oint(\nb\circ\nb)\dd S.
\end{equation}
In CSL, the c.o.m. decoherence of  homogeneous bulks is  a \emph{surface effect}!

Observe that the  main result (\ref{Deccomxxgeom}) remains valid if the probe has cavities in it. 
This allows us to multiply the CSL decoherence
By carving cavities inside the otherwise homogeneous probe,  CSL decoherence can be multipled (cf. Fig. \ref{fig-1}).
This explains the reason of enhanced decoherence in layered structures, proposed by \cite{CarVinBas18}.

The heating rate $\Gamma_\cm =\Dcal_\cm (\Pbh^2/2M)$
of the c.o.m. motion is now easy to write in a more
explicite form than before. Reading $\Dcal_\cm $ off from (\ref{Deccomxxgeom}), one immediately obtains
\begin{equation}\label{heatcm}
\Gamma_\cm =\frac{2\pi\lambda\si^2\varrho^2}{m_N^2}\frac{S}{M}
=\frac{2\pi\lambda\si^2\varrho}{m_N^2}\frac{S}{V},
\end{equation}
where $S$ is the total surface (including cavities' internal surfaces) and $V$ is the total volume (excluding cavities). 
Note that $\Gamma_\cm $ is the same if we start from the general dynamics (\ref{Deccom}) not restricted by $\vert\Delta\Xb\vert\ll\si$.
Interestingly, c.o.m. heating is inverse proportional
to the size of the bulk. Recall the total heating rate
\begin{equation}
\Gamma=\Dcal\sum_a\frac{\pbh^2}{2m_a}=\frac{3\hbar^2\lambda}{2m_N^2\si^2}M,
\end{equation}
always much larger than the c.o.m. heating. For a sphere of radius $R$ we get 
$\Gamma_\cm/\Gamma=3(\si/R)^4$. 

\emph{Examples.}
Consider the longitudinal motion of a cylinder, Eq. (\ref{Deccomxxgeom}) reduces to
\begin{equation}\label{Deccomxxgeomrod}
\Dcal_\cm \ro_\cm =-\frac{2\pi\lambda\si^2\varrho^2}{m_N^2}S_\perp[\hat x,[\hat x,\ro_\cm ]],
\end{equation}
where $S_\perp$ is the total surface perpendicular to the motion
(i.e.: the area of both faces of the cylinder). 
At a given constant density $\varrho$, the decoherence is independent of the
length of the cylinder. It can be squeezed to become a plate or elongated to become
a rod. This invariance of the decoherence
offers a fair guidance when we design laboratory probes. However, the
same invariance may raise conceptual questions as well. With increasing
length of the rod while decoherence rate remains constant, 
CSL might leave the longitudinal superposition of our massive rod with 
counter-intuitive long coherence times. 
An other remarkable feature of the surface-tensor $\mathrm{S}$ is
that spontaneous decoherence in one direction can be decreased by tilted
edges instead of perpendicular ones. If the faces of the cylinder 
are replaced by cones of apex angle $\theta$ then 
spontaneous longitudinal decoherence becomes suppressed by the
factor $\sin(\theta/2)$. E.g.: sharp pointed needles become extreme 
insensitive to longitudinal CSL.  

\section{Rotational decoherence}\label{IV}
Our main result (\ref{Deccomxxgeom}) on decoherence of lateral superpositions
tells us how to calculate decoherence of angular superpositions.
It turns out that rotational decoherence, too,  is a surface effect. 
Let us consider small rotations around a single axis $\nb_\rt$ for convenience.  
The small lateral displacement $\nb\Xbh-\langle\nb\Xbh\rangle$ ---effective in (\ref{Deccomxxgeom})---
will be replaced by the small rotational displacement
$\nb(\rb\times\nb_\rt)(\hat\varphi-\langle\hat\varphi\rangle)$
where $\hat\varphi$ is the angle of rotation.
Then, with the scalar triple product notation,  the main equation  (\ref{Deccomxxgeom}) reads:
\begin{equation}\label{Deccomphiphigeom}
\Dcal_\cm ^\rt\ro_\cm =-\frac{2\pi\lambda\si^2\varrho^2}{m_N^2}
\oint [\rb,\nb,\nb_\rt]^2 \dd S~[\hat\varphi,[\hat\varphi,\ro_\cm ]].
\end{equation}
Rotational decoherence is determined by the
constant density $\varrho$ and the \emph{rotational surface-tensor}: 
\begin{equation}
\mathrm{S}_\rt=:\oint(\rb\times\nb)\circ(\rb\times\nb)\dd S.
\end{equation}
Remember, our starting equation (\ref{Deccomxxgeom}) was valid for $\vert\Delta \Xb\vert\ll\si$ only, hence
the validity of (\ref{Deccomphiphigeom}) requests the corresponding smallness of the angular uncertainties. 

Calculation of the spontaneous heating rate of the rotational degrees of freedom is straightforward, yielding
\begin{equation}\label{heatrot}
\Gamma_\rt =\frac{2\pi\lambda\si^2\varrho}{m_N^2}\mathrm{Tr}(\mathrm{I}^{-1}\mathrm{S}_\rt),
\end{equation}
where $\mathrm{I} =\int (\rb\circ\rb)\dd\rb$ is the inertia tensor of the probe.

\emph{Examples.} 
Consider the rotation of a long cylindric rod of length $L$ and radius $R\ll L$, around a perpendicular axis
through its center. 
All along the rod ---except for its short middle part of size $\sim\!R$--- the expression
$[\rb,\nb,\nb_\rt]=r\sin(\Phi)$ is a good approximation where
$r\in(-L/2,L/2)$ is the axial coordinate and $\Phi$ is the azimuthal angle of the surface position $\rb$. 
Using this approximation, we can easily evaluate the axial element of the rotational surface-tensor $\mathrm{S}_\rt$
that controls the angular decoherence (\ref{Deccomphiphigeom}): 
\begin{equation}
\oint [\rb,\nb,\nb_\rt]^2 \dd f =  \frac{\pi RL^3}{12}.
\end{equation} 
As another example, consider  our cylinder rotating around its axis of symmetry: CSL predicts
zero decoherence (cf. Fig. \ref{fig-1}). But we introduce a small elliptical excentricity $e\ll1$ of
the cross section.
In leading order, we have $[\rb,\nb,\nb_\rt]=\frac{1}{2} R e^2\sin(2\Phi)$, yielding
the following contribution of the shape to the strength of angular decoherence:
\begin{equation}
\oint [\rb,\nb,\nb_\rt]^2 \dd f =  \frac{e^4}{4}\pi R^2 L,
\end{equation} 
that is $e^4/4$ times the volume of the cylinder. Recall that $e^2=2\Delta R/R$
where $\Delta R$ is the small difference between the main diameters of
the elliptic cross section. The obtained result may raise the same conceptual problem
that we mentioned for the longitudinal superposition of the massive rod/needle:
azimuthal superpositions of massive cylinders of low excentricity may
become practically insensitive to CSL.

\section{Outlines of generalizations}\label{V}
That in CSL the c.o.m and rotational decoherences are surface effects for homogeneous probes has been
explicitly shown in Secs. \ref{III} and \ref{IV} for ideal sharp edges and for spatial superpositions much smaller
than $\si$. Both of the latter restrictions can be relaxed and $\Dcal_\cm $ still remains a surface integral.

The case of unsharp edges is not much different from the ideal case. If the profile $H(h)\varrho$ of how the 
density drops from the constant $\varrho$ down to zero through a thin layer defining the surface where 
the layer's thickness is small w.r.t. the sizes of the probe then the following generalization of Eq. (\ref{profile0}) helps:
\begin{equation}\label{profile1}
\nabla\mu_\si(\rb+h\nb)=\varrho \nb \int g_\si(h-h^\prime)\dd H(h^\prime).
\end{equation}
The rest of constructing the surface integral is the same as for Eq. (\ref{profile0}) which described
the special case where $H$ was the (descending) step function.

The case of not necessarily small quantum positional and angular quantum uncertainties was described by
Eq. (\ref{Deccom}). It takes an equivalent closed form in coordinate representation:
\begin{equation}\label{Deccomcoord}
\Dcal_\cm \ro_\cm (\Xb,\Yb)=-\frac{\lambda\si^3}{\pi^{3/2}m_N^2}
(2\pi)^3\int\left[\mu_\si(\rb+\Xb)\mu_\si(\rb+\Yb)-\mu_\si^2(\rb)\right]\dd\rb~\ro_\cm (\Xb,\Yb).
\end{equation}
The relevant structure is the integral, which we write as
\begin{equation}
(2\pi)^3\int\left[\mu_\si(\rb+\Xb-\Yb)-\mu_\si(\rb)\right]\mu_\si(\rb)\dd\rb.
\end{equation}
As long as the quantum uncertainty $\vert\Xb-\Yb\vert$ is much smaller than the sizes of the probe,
the integral is vanishing everywhere in the bulk except for a thin layer of thickness $\sim\!\vert\Xb-\Yb\vert$
around the surface. Accordingly, CSL decoherence remains a surface effect and, investing some harder
mathematical work, $\Dcal_\cm$ as well as $\Dcal_\rt$ would take a form of surface integral, generalizing 
(\ref{Deccomxxgeom}) and (\ref{Deccomphiphigeom})
beyond their quadratic approximations in $\Xbh$ and $\hat\varphi$.

\section{Concluding remarks}\label{VI}
We have discussed CSL for constant density test masses and proved that spontaneous decoherence 
of both translational and
rotational motion is determined by the density $\varrho$ and by two invariant surface-tensors of the bodies: 
$$
\mathrm{S}=\oint(\nb\circ\nb)\dd S,
$$
$$
\mathrm{S}_\rt=\oint(\rb\times\nb)\circ(\rb\times\nb)\dd S.
$$
These two fully encode the relevant features of the probe's geometry.   
Previously, these features were encoded by the so-called geometric factor
$$
\mu_\kb=\varrho\int \erm^{-\im\kb\rb}\dd\rb,
$$
an integral over the probe's volume and a function of the wave number $\kb$. 
In case of general heavily inhomogeneous test masses the necessity of using the geometric 
factor is certainly doubtless. But for homogeneous probes, the surface-tensors 
should take over the role.

Important is the new insight into the physics of CSL in motion
of a general massive bulk as a whole. First, microscopic structure is totally
irrelevant, only the $\si$-smoothened density matters. Furthermore,
displacements of homogeneous regions are not decohered at all. Only
the displacements of inhomogeneities are decohered.  The sharper the
inhomogenity, the stronger the decoherence it induces.  In a constant
density probe, the only inhomogeneous part is its surface, hence
is CSL decoherence a surface effect for it --- that we have here exploited.
But surface inhomogeneity is a sharpest possible one, and decoherence 
for probes with smooth inside inhomogeneities is likely to remain 
dominated by the surface, our method of surface-tensors might 
remain valid for them!
Layer inhomogeneities with thin walls between them
are competitive, their effect is surface effect and our surface-tensors
could be generalized to include them. Whether and when 
lower than two-dimensional inhomogeneities could play a role ---
that worth investigation.
\vskip1cm

The author thanks the National Research Development and Innovation Office of Hungary Projects
Nos. 2017-1.2.1-NKP-2017-00001 and K12435, and the EU COST Action CA15220 for support.

\end{document}